\documentclass[11pt,a4paper]{article}
\usepackage{graphicx}
\usepackage{jcappub}
\usepackage{lscape}

\newcommand{\simgt}{\lower.5ex\hbox{$\; \buildrel > \over \sim \;$}}
\newcommand{\simlt}{\lower.5ex\hbox{$\; \buildrel < \over \sim \;$}}

\makeatletter
\newsavebox{\@parc@ption}
\def\parcaption#1{%
\sbox{\@parc@ption}{\shortstack[l]{#1}}%
\setbox\@tempboxa\hbox{\csname fnum@\@captype\endcsname}%
\@tempdima\columnwidth \advance\@tempdima-\wd\@tempboxa
\@tempdimb\@tempdima 
\ifdim\wd\@parc@ption>\@tempdimb \@tempdima\@tempdimb
\else\@tempdima\wd\@parc@ption\fi
\sbox{\@tempboxa}{\parbox[t]{\@tempdima}{#1}}%
\caption{\usebox{\@tempboxa}}}
\makeatother

\title{  Testing a generalized cubic Galileon gravity model with the Coma Cluster
  }

\author{
  Ayumu Terukina${}^1$, Kazuhiro Yamamoto${}^{1,2}$, Nobuhiro Okabe${}^{1,2}$, 
  Kyoko Matsushita${}^3$ and  Toru Sasaki${}^3$
}

\affiliation{
${}^1$Department of Physical Sciences, Hiroshima University, 
1-3-1 Kagamiyama, Higashi-Hiroshima, Hiroshima 739--8526, Japan
\\
${}^2$Hiroshima Astrophysical Science Center, Hiroshima University, 1-3-1 Kagamiyama, Higashi-Hiroshima,
Hiroshima 739--8526, Japan
\\
${}^3$Department of Physics, Tokyo University of Science, 1-3 Kagurazaka,
Shinjuku-ku, Tokyo 162--8601, Japan
}

\abstract{

We obtain a constraint on the parameters of a generalized cubic Galileon 
gravity model exhibiting the Vainshtein mechanism by using multi-wavelength 
observations of the Coma Cluster.
The generalized cubic Galileon model is characterized by three
parameters of the turning scale associated with the Vainshtein
mechanism, and the amplitude of modifying a gravitational potential
and a lensing potential.
X-ray and Sunyaev--Zel'dovich (SZ) observations of the
intra-cluster medium are sensitive to the gravitational potential, while 
the weak-lensing (WL) measurement is specified by the lensing potential. 
A joint fit of a complementary multi-wavelength dataset of X-ray, SZ and WL
measurements enables us
to simultaneously constrain these three parameters of the generalized
cubic Galileon model for the first time.
We also find a degeneracy between the cluster mass parameters
and the gravitational modification parameters, which is influential 
in the limit of the weak screening of the fifth force.
}
\arxivnumber{}

\begin{document}
\maketitle

\section{Introduction} 
\label{sec:intro}
Modifications of gravity theory is an interesting approach to 
explaining the accelerated expansion of the universe.
However, any covariant modification of general relativity introduces 
additional degrees of freedom, giving rise to a fifth force.
This is strictly constrained by gravity tests in the solar system.
Solar system experiments \cite{Adelberger,Will}
  are in excellent agreement with
general relativity, requiring that this additional degree of freedom 
be hidden on the scale of the solar system.
Such a process is referred to as a ``screening mechanism,'' which is key for a viable modified gravity model.
In general, this screening mechanism works in high-density 
regions where the matter density contrast is nonlinear.  
However, this does not work on large cosmological scales. 
A screening mechanism that characterizes viable modified gravity 
models is an important feature to be tested with observations.

The chameleon mechanism \cite{Khoury,Mota} is a screening mechanism 
that works in an $f(R)$ gravity model and the chameleon gravity model 
\cite{Starobinski,HS,Tsujikawa}. In these models, 
a scalar degree of freedom that gives rise to the fifth force is screened
in a high-density region due to coupling with matter. The 
chameleon gravity model and an $f(R$) model can be viable owing to the chameleon mechanism \cite{Lucas}. 
The Vainshtein mechanism \cite{Vainshtein} is another relevant screening mechanism,
which is exhibited in the Dvali--Gabadaze--Porrati (DGP) model \cite{DGP,Deffayet1}, 
the simplest cubic Galileon model (\cite{Nicolis,Deffayet3,Deffayet2,Luty}, see also the appendix),
and its generalized version \cite{Kobayashi,Kimura}. 
The DGP model is an archetypal modified gravity model developed in the 
context of the brane-world scenario. There are two branches of solutions in the DGP model. 
The self-acceleration
branch DGP (sDGP) model \cite{Koyama1,Koyama2,Schmidt1} includes a mechanism to explain self-acceleration in the late 
universe, while the normal branch DGP (nDGP) model \cite{Deffayet4,Falck,Schmidt2} with a cosmological constant 
is a healthy modified gravity model avoiding the ghost problem \cite{Nicolis2,Goubnov}. 
The simplest cubic Galileon model is also a typical modified gravity model that explains 
self-acceleration of the universe while avoiding the ghost problem. 
Our generalized cubic Galileon model is a generalized version of the simplest cubic Galileon 
model that retains important features and contains the DGP models.
In these models, a scalar field giving rise to a fifth force is screened
due to self-interaction on small scales where density 
perturbations become nonlinear. 

Galaxy clusters provide a unique laboratory
for testing modified gravity models 
exhibiting screening mechanisms, because they are objects on the borderline 
between linear and nonlinear scales, that is, 
between non-screened and screened scales. 
The authors of \cite{Terukina1,Terukina2,Harry} have investigated a cosmological 
constraint on the chameleon gravity model using galaxy clusters. 
They put a constraint on the turning scale and the amplitude of the fifth force
at large scales. 
The authors of \cite{Narikawa,Narikawa2} have investigated a constraint on 
a generalized Galileon model exhibiting the Vainshtein mechanism, using an 
observed weak-lensing profile of clusters.
They put a constraint on the turning scale and the amplitude of 
modification of the lensing potential. 

The purpose of the present paper is twofold. One purpose is a generalization of 
the methodology for testing a modified gravity model with a galaxy cluster. 
In the present paper, we consider a generalized cubic Galileon model. 
Within the quasi-static approximation, the generalized cubic Galileon model
is effectively characterized by three parameters, $\mu_{\rm G}$, $\mu_{\rm L}$ and $\epsilon$.  
Detailed definitions are given later, but, broadly,
$\mu_{\rm G}$ and $\mu_{\rm L}$ are parameters that modify the effective amplitude of the
gravitational potential and the lensing potential in the non-screened region, 
while $\epsilon$ determines the turning scale from the non-screened region
to the screened region due to the Vainshtein mechanism. 
The parameters $\mu_{\rm G}$ are constrained by observations of the gas distribution, 
in particular an X-ray surface brightness profile and the SZ effect \cite{SZ}. 
However, the parameter $\mu_{\rm L}$ is only constrained by observations of lensing 
measurements. 
Therefore, a combination of observations of gas distribution and the 
lensing signal is essential to put a constraint on the three parameters 
characterizing the modified gravity model. 
We demonstrate how a combination of multi-wavelength observations of a 
cluster is useful to put a constraint on a generalized Galileon model. 

The other purpose is improvement of the analysis in \cite{Terukina2} using 
new X-ray data \cite{Coma_xray1,Coma_xray2} and lensing \cite{Coma_WL2} observations of 
the Coma Cluster. 
In our method of testing gravity with a galaxy cluster, 
the modeling of the gas distribution is important.
A basic assumption of the model for the gas distribution is 
hydrostatic equilibrium, that is, 
a balance between the gas pressure gradient force and the gravitational force. 
In the region where the fifth force is influential, the condition of the 
hydrostatic equilibrium is changed, and the gas density profile is modified. 
However, in general, galaxy clusters are dynamically evolving, and a deviation 
from the equilibrium could be influential. Therefore, we first check the 
consistency of our model by comparing theoretical predictions with
various observations of the Coma Cluster, including the new X-ray data and 
lensing measurements. 

This paper is organized as follows: In Section~\ref{sec:modeling}, we first review our model 
for the dark matter and gas distribution of a galaxy cluster within 
Newtonian gravity. We demonstrate how well our model fits observations 
of the Coma Cluster in Section~\ref{sec:Coma}. We also validate our model against the influence of non-thermal 
pressure. 
In Section~\ref{sec:GG}, we introduce a generalized cubic Galileon model, and explain the 
modification of our model by the fifth force of the scalar field. 
In Section~\ref{sec:Discussion}, we discuss degeneracies on parameters and systematic errors focusing on 
special circumstance of using the Coma Cluster.
Section~\ref{sec:summary} is devoted to a summary and conclusions. 
Throughout this paper, we adopt $\Omega_{\rm m0}=0.3$, $\Omega_\Lambda=0.7$ and
$H_0=70{\rm km/s/Mpc}$, and we follow the convention $(-,+,+,+)$.

\section{Modeling of cluster profiles} 
\label{sec:modeling}
We review our model for the dark matter and gas density 
distribution connecting with observational quantities,
X-ray brightness,
 the SZ effect temperature profile, and the weak-lensing 
profile (see also \cite{Terukina2,Narikawa}). Our model is based on an 
assumption of hydrostatic equilibrium, but we also take the possible 
influence of non-thermal pressure 
into account. In this section, we first consider the case of Newtonian gravity. 

\subsection{Mass profile} 
\label{sec:mass}
We employ a universal mass density profile, namely the Navarro--Frenk--White (NFW) 
profile \cite{NFW}, motivated by predictions of numerical simulations,
\begin{align}\label{NFW}
\rho(r)=\frac{\rho_{\rm s}}{r/r_{\rm s}(1+r/r_{\rm s})^2},
\end{align}
where $\rho_{\rm s}$ and $r_{\rm s}$ are the normalization and the scale
radius, respectively.
The mass 
within the radius $r$ is given by $
M(<r)=4\pi\int^r_0drr^2\rho(r)=4\pi\rho_{\rm s}r_{\rm s}^3m(r),
$ with $m(r)=\ln(1+r/r_{\rm s})-r/r_{\rm s}/(1+r/r_{\rm s})$.
We introduce the concentration parameter $c$ and the virial mass $M_{\rm vir}$
instead of $\rho_{\rm s}$ and $r_{\rm s}$,
\begin{align}
&c= \frac{r_{\rm vir}}{r_{\rm s}},\\
&M_{\rm vir}\equiv M(<r_{\rm vir})=\frac{4\pi}{3}r^3_{\rm vir}\Delta_{\rm c}\rho_{\rm c}(z)
\end{align}
with the virial radius $r_{\rm vir}$.
Here, $\rho_{\rm c}$ is the critical density at the redshift $z$,
and $\Delta_{\rm c}= 100$ is the overdensity contrast determined by the
spherical collapse model \cite{Nakamura}.

\subsection{Gas profile with hydrostatic equilibrium} 
\label{sec:gas}
We first assume hydrostatic equilibrium between the gas pressure 
gradient and the gravitational force in
galaxy clusters as
\begin{align}\label{hydroeq}
\frac{1}{\rho_{\rm gas}}\frac{dP_{\rm tot}}{dr}=-\frac{d\Psi}{dr},
\end{align}
where $\rho_{\rm gas}$ is the gas density, $P_{\rm tot}=P_{\rm th}+P_{\rm nth}$ is the sum of the
thermal pressure, $P_{\rm th}$ and the non-thermal pressure, $P_{\rm nth}$ (see below for details),
and $\Psi$ is the gravitational potential.
We employ three assumptions to describe the gas physics. 
First, the equation of state for the gas components, $P_{\rm th}=n_{\rm
gas}kT_{\rm gas}$, where $n_{\rm gas}$ and $T_{\rm gas}$ are 
the number density and the temperature of the total gas component,
respectively. Second, the temperature of electrons is 
the same as that of the gas, $T_{\rm e}=T_{\rm gas}$.
Third, the electron pressure satisfies $P_{\rm e}=n_{\rm e}kT_{\rm gas}$ 
with the electron number density $n_{\rm e}=(2+\mu)n_{\rm gas}/5$, 
where $\mu=0.59$ is the mean molecular weight.
In the present paper, for the electron temperature we assume the
functional form
\begin{align}\label{temperature}
T_{\rm e}(r)=T_0\left[1+\left(\frac{r}{r_{\rm 1}}\right)^{b_1}\right]^{-b_2/b_1},
\end{align}
where $T_0$, $b_1$, $b_2$ and $r_1$ are parameters.
Integrating (\ref{hydroeq}) with (\ref{temperature}), we obtain
the electron pressure profile
\begin{align}\label{pressureB1}
P_{\rm e}(r)=n_0T_0\exp\left(\int_0^rdr \frac{\mu m_{\rm p}}{kT_{\rm e}(r)}
\left[-\frac{GM(<r)}{r^2}\right]\right),
\end{align}
and the electron number density $n_{\rm e}(r)=P_{\rm e}(r)/kT_{\rm e}(r)$,
where $n_0$ is the normalization parameter of the electron number density, $n_{\rm e}$.
In deriving equation (\ref{pressureB1}),
we use the relation, $\rho_{\rm gas}=\mu m_{\rm p}n_{\rm gas}$, where $m_{\rm p}$ is the proton mass.
Equation (\ref{pressureB1}) is the case of the absence of non-thermal 
pressure; the case including non-thermal pressure is described below. 

Thus our gas distribution model includes 7 parameters
($M_{\rm vir}$,$c$,$n_{0}$,$T_0$,$b_1$,$b_2$,$r_1$).
Using our model of the three-dimensional profiles, we construct 
the observables for the observations of X-ray and the cosmic microwave background (CMB) temperature distortion.
The X-ray emission from clusters are dominated by the bremsstrahlung and line emission caused by
the ionized gas. For the X-ray observable, we define the X-ray brightness as $B_{\rm X}\equiv {\it norm}/{\it area}$,
where {\it norm} is the spectrum normalization obtained from XSPEC software \cite{XSPEC}
adopting the APEC emission spectrum \cite{APEC}, and {\it area} is the area of the spectrum.
The spectrum normalization is given by ${\it norm} \propto \int n_{\rm e}n_{\rm H}dV$, where
$n_{\rm H}=0.86n_{\rm e}$ is the hydrogen number density and $V$ is the volume of the spectrum.
Then, we write the X-ray brightness as
\begin{align}
  B_{\rm X}(r_\perp)=\frac{10^{-14}}{4\pi (1+z)^2}\int n_{\rm e}(r)n_{\rm H}(r)dz ~[{\rm cm^{-5}/arcmin^2}],
\end{align}
where $r_\perp$ is the radius perpendicular to the line-of-sight
direction, which is related with $r$ and $z$ as $r=\sqrt{r_\perp^2+z^2}$.
The CMB temperature distortion is caused by CMB photons passsing through clusters and being scttered
by electrons in clusters, can be expressed as the difference between the averaged CMB temperature
and the observed CMB temperature, $\Delta T_{\rm SZ}$, or $y$-parameter,
\begin{align}
  y(r_\perp)=-\frac{\Delta T_{\rm SZ}}{2T_{\rm CMB}}=\frac{\sigma_{\rm T}}{m_{\rm e}}\int P_{\rm e}(r)dz,
\end{align}
where $ T_{\rm CMB}=2.725$ K is the CMB temperature, $\sigma_{\rm T}$ is the Thomson cross section, $m_{\rm e}$ is
the electron mass.

\subsection{Shear profile by gravitational weak-lensing} 
\label{sec:lensing}
We consider a spatially flat cosmological background, and work 
with the cosmological Newtonian gauge, whose line element is written as
\begin{eqnarray}
ds^2=-(1+2\Psi)dt^2+a(t)^2(1+2\Phi)d{\bf x}^2,
\label{metric}
\end{eqnarray}
where $a(t)$ is the scale factor, and $\Psi$ and $\Phi$ are the gravitational and curvature 
potentials, respectively.
The propagation of light is determined
by the lensing potential $(\Phi-\Psi)/2$, which means that the weak-lensing signal is 
determined by $(\Phi-\Psi)/2$. 
For example, the convergence is given by
\begin{align}\label{kappa1}
\kappa= -\frac{1}{2}\int_0^{\chi}d\chi^\prime\frac{(\chi-\chi^\prime)
\chi^\prime}{\chi}\triangle^{\rm (2D)}(\Phi-\Psi),
\end{align}
where $\chi$ is the comoving distance and 
$\triangle^{\rm (2D)}$ is the comoving two-dimensional Laplacian.
For the case of general relativity, we may set 
$\triangle\Psi=-\triangle\Phi=4\pi Ga^2\rho$. Then, using the thin lens approximation, 
{(\ref{kappa1})} reduces to 
\begin{align}\label{kappa2}
\kappa=\frac{(\chi_{\rm S}-\chi_{\rm L})\chi_{\rm L}}
{\chi_{\rm S}}
\int_0^{\chi_{\rm S}}d\chi'\left[4\pi G\rho(r^\prime)\right]a_{\rm L}^2,
\end{align}
where $\chi_{\rm L}$ and $\chi_{\rm S}$ denote the comoving distance between the 
observer and lens and that between the observer and the source, respectively, 
and $a_{\rm L}=1/(1+z_{\rm L})$ is the scale factor specified by the 
redshift of the lens object $z_{\rm L}$. 
For a spherically symmetric cluster, (\ref{kappa2}) is represented as
\begin{align}\label{kappa3}
\kappa(r_\perp)=
\frac{2}{\Sigma_{\rm c}}
\int_0^{\infty}dz\rho(r)
\end{align}
with the physical coordinate $r=\sqrt{r_\perp^2+z^2}$. We define
$\Sigma_{\rm c}=
\chi_{\rm S}/[{4\pi G}
{(\chi_{\rm S}-\chi_{\rm L})\chi_{\rm L}a_{\rm L}}]
$.
We then define the reduced shear
\begin{align}\label{reduced_shear}
g_+(r_\perp)\equiv\frac{\gamma_+(r_\perp)}{1-\kappa(r_\perp)},
\end{align}
where $\gamma_+(r_\perp)$ is the tangential shear, which is related with the 
convergence as
\begin{align}\label{reduced_shear}
\gamma_+(r_\perp)=\bar\kappa(<r_\perp)-\kappa(r_\perp),
\end{align}
with
\begin{align}\label{barconvergence}
\bar\kappa(<r_\perp)\equiv \frac{2}{r_\perp^2}\int_0^{r_\perp}dr_\perp^\prime r_\perp^\prime
\kappa(r_\perp^\prime).
\end{align}
For the NFW profile, the convergence is given by \cite{NFWL} as
\begin{align}
  \kappa_{\rm nfw}(x)=\left\{
  \begin{array}{ll}
    \dfrac{2r_{\rm s}\rho_{\rm s}}{\Sigma_{\rm c}(x^2-1)}\left[1-\dfrac{2}{\sqrt{1-x^2}}
      \text{arctanh}\sqrt{\dfrac{1-x}{1+x}}\right], &\quad (x<1) \\
    \dfrac{2r_{\rm s}\rho_{\rm s}}{3\Sigma_{\rm c}}, &\quad (x=1)\\
    \dfrac{2r_{\rm s}\rho_{\rm s}}{\Sigma_{\rm c}(x^2-1)}\left[1-\dfrac{2}{\sqrt{x^2-1}}
      \text{arctan}\sqrt{\dfrac{x-1}{1+x}}\right], &\quad (x>1)
  \end{array}
  \right.
\end{align}
\begin{align}
  \bar\kappa_{\rm nfw}(<x)=\left\{
  \begin{array}{ll}
    \dfrac{4r_{\rm s}\rho_{\rm s}}{\Sigma_{\rm c}x^2}\left[\dfrac{2}{\sqrt{1-x^2}}
      \text{arctanh}\sqrt{\dfrac{1-x}{1+x}}+\ln\left(\dfrac{x}{2}\right)\right], &\quad (x<1) \\
    \dfrac{4r_{\rm s}\rho_{\rm s}}{\Sigma_{\rm c}}\left[1+\ln\left(\dfrac{1}{2}\right)\right], &\quad (x=1)\\
    \dfrac{4r_{\rm s}\rho_{\rm s}}{\Sigma_{\rm c}x^2}\left[\dfrac{2}{\sqrt{x^2-1}}
      \text{arctan}\sqrt{\dfrac{x-1}{1+x}}+\ln\left(\dfrac{x}{2}\right)\right], &\quad (x>1)
  \end{array}
  \right.
\end{align}
with $x=r/r_{\rm s}$.

Here, we assume that the source galaxies have random orientation of ellipticity 
$\epsilon_{\rm S}$, the average of which is $\langle\epsilon_{\rm S}\rangle=0$.
When we observe the tangential ellipticity of the source galaxies 
$\epsilon_{\rm obs.}$, the average is given by $\langle \epsilon_{\rm obs.}\rangle= g_+$.
Hereafter, we assume that the redshift of the source galaxies
is 
$\langle z_{\rm S}\rangle=0.6$, but the results are not influenced 
by the redshift of the source galaxies for nearby clusters.

\section{Consistency test with Newtonian gravity} 
\label{sec:Coma}
In the present paper, we use Coma Cluster observations.
The Coma Cluster is one of the best observed nearby clusters, and has redshift
$z=0.0236$. The X-ray distribution
\cite{Coma_xray1,Coma_xray2,Coma_xray3,Coma_xray4,Coma_xray5,Coma_xray6,Coma_SB,Coma_XT1,Coma_XT2,Gastaldello},
the SZ effect \cite{Coma_SZ} and the weak-lensing measurement \cite{Coma_WL1,Coma_WL2} have been reported.
These observations revealed that the Coma Cluster has
substructures and orientation dependence on the gas temperature profiles.
The Coma Cluster is thus an unrelaxed system. However, we will show that 
our model based on the hydrostatic equilibrium fits the data of 
the X-ray brightness profiles  \cite{Coma_xray1,Coma_xray2}, 
the SZ effect profile from the Planck measurement \cite{Coma_SZ}, 
and the weak-lensing profile by Subaru observations \cite{Coma_WL2}.
In general, the assumption of hydrostatic equilibrium holds
  only at the intermediate region of clusters,
  because of the cooling of the gas at the innermost region and the environmental effects
  at the outermost region. Then we use data points in the range 
  $200\ {\rm kpc}$ to $1.5\ {\rm Mpc}$
  to get rid of systematic effects from the innermost and outermost 
regions of the cluster.

In this work, we use the observational data of the XMM-Newton 
\cite{Coma_xray1,Coma_xray2}, which are different from those 
used in a previous paper \cite{Terukina2}. 
In that paper, the weak-lensing profile
is not used; only the parameters $M_{\rm vir}$ and $c$ are used as a 
prior profile from \cite{Coma_WL1}. However, use of the weak-lensing profile is essential 
to our analysis of the generalized Galileon model. 

To address the theoretical predictions in the
previous section with observations of the Coma Cluster,
we introduce the chi-squared by summing the chi-squared for each observation as
\begin{align}
\label{chi2}
\chi_{\rm XB+SZ+WL}^2&=
\chi_{\rm XB}^2+\chi_{\rm SZ}^2+\chi_{\rm WL}^2,
\end{align}
where 
\begin{align}
&\chi_{\rm XB}^2=
\sum_i\frac{(B_{\rm X}(r_{i})-B_{{\rm X},i}^{\rm obs.})^2}{(\Delta B_{{\rm X},i}^{\rm obs.})^2},\\
&\chi_{\rm SZ}^2=
\sum_i\frac{(y(r_{\perp,i})-y_{i}^{\rm obs.})^2}{(\Delta y_{,i}^{\rm obs.})^2},\\
&\chi_{\rm WL}^2=
\sum_i\frac{(g_{+}(r_{\perp,i})-g_{+,i}^{\rm obs.})^2}{(\Delta g_{+,i}^{\rm obs.})^2},
\end{align}
are the chi-square values for the X-ray brightness, the SZ effect and the weak-lensing, respectively.
We note that the covariance of errors is not taken into account in our analysis
and leave it for future work to study how the observational systematics affect our analysis.

\begin{landscape}
\begin{figure}[b]
\centering
\includegraphics[width=1\hsize]{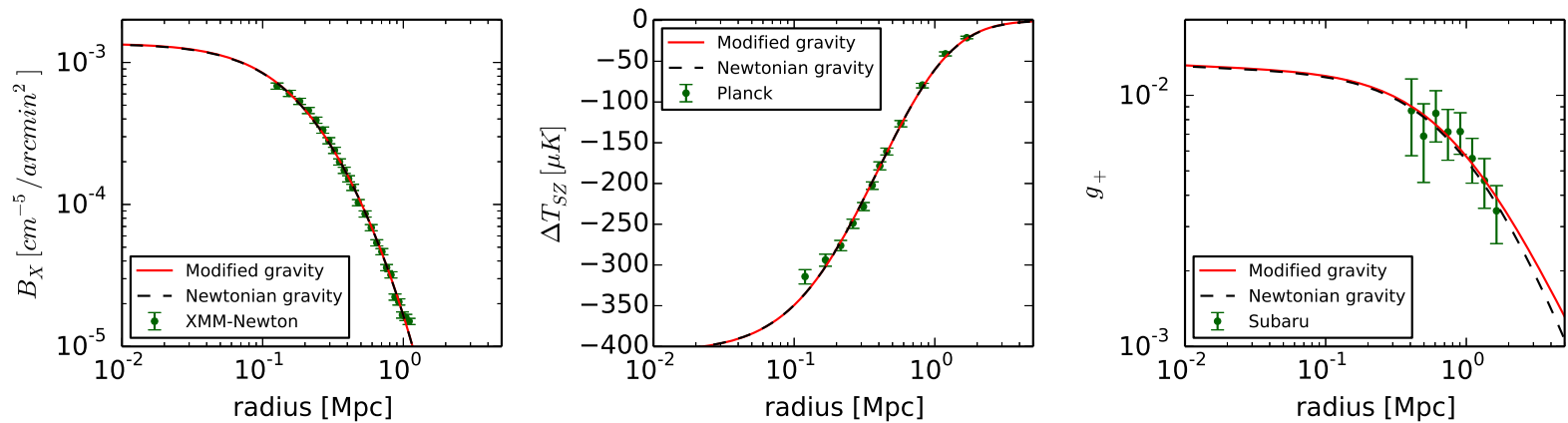}
\caption{
Best-fit profiles in the frameworks of Newtonian gravity (black dashed curve) 
and the generalized Galileon model (red solid curve),  and the observational results. 
The best-fit parameters are listed in Table~\ref{table1}.
\emph{Left panel}:
The X-ray surface brightness from the {\it XMM-Newton} observations 
\cite{Coma_xray1,Coma_xray2}.
The errors bars are composed of the Poisson noise and systematic
 errors that we here assume as 5\%.
\emph{Center panel}:
The SZ temperature profile from the {\it Planck} measurements \citep{Coma_SZ}.
\emph{Right panel}: The weak-lensing profile from the Subaru observations \cite{Coma_WL1}.
    \label{fig:observed2_MG}
  }
\end{figure}
\end{landscape}

We perform an MCMC analysis using modified {\it Monte Python} code \cite{MP}
that employs a Metropolis--Hastings \cite{Metropolis,Hastings} sampling
algorithm. This analysis includes 5 parameters with the chi-squared
(\ref{chi2}), $\chi^2_{\rm BX+SZ+WL}$.
We require Gelman--Rubin statistics \cite{Gelman} of ${\cal R}-1<0.001$
for each parameter to ensure convergence of our runs.
The black dashed curve in each panel of figure~\ref{fig:observed2_MG} shows the best-fit profiles
for the Newtonian gravity.
The minimum value of the chi-squared is $\chi_{\rm XB+SZ+WL}^2/{\rm d.o.f.}=58/44$,
and the 2-dimensional marginalized contours of the different combinations between the
 model parameters are shown in figure~\ref{fig:mcmc0}.

\begin{table}[t]
\caption{
  Best-fitting parameters and 1-dimensional marginalized constraints ($68$\% CL)
  to characterize the gas and lensing profiles.
  To avoid degeneracy between parameters,  we fix $b_1$ and $b_2$ simply.
  Our results do not depend on whether these parameters are fixed or not.
  This table shows the results for the Newtonian gravity (second column) and
    the generalized Galileon model with all modification parameters (third column).
    The minimum chi-squared and the number of degrees of freedom,
    d.o.f. $=$ (number of data points) $-$ (number of model parameters), are listed at the bottom
    of each column.
\label{table1}
}
\begin{center}
  \renewcommand{\arraystretch}{1.5}
\begin{tabular}{c|c|c}
\hline\hline
Parameter & ~~~~Newtonian gravity~~~~ & Modified gravity (full parameters)
\\
\hline
$M_{\rm vir}$  & ${1.08}^{+0.06}_{-0.06}\times 10^{15}~M_\odot$    & ${1.04}^{+0.14}_{-0.06}\times 10^{15}~M_\odot$ \\
$c$           & ${3.59}^{+0.23}_{-0.23}$                         & ${3.64}^{+0.21}_{-0.30}$ \\
$n_0$         & ${6.14}^{+0.28}_{-0.26}\times 10^{-3}{\rm /cm^3}$ & ${6.17}^{+0.26}_{-0.31}\times 10^{-3}{\rm /cm^3}$ \\
$T_0$         & ${6.36}^{+0.11}_{-0.12} ~{\rm keV}$              & ${6.35}^{+0.13}_{-0.11} ~{\rm keV}$ \\
$b_1$         & ${2.6}$ (fixed)                                & ${2.6}$ (fixed) \\
$b_2$         & ${0.5}$ (fixed)                                & ${0.5}$ (fixed) \\
$r_1$         & ${0.74}^{+0.06}_{-0.06} ~{\rm Mpc}$              & ${0.75}^{+0.06}_{-0.07} ~{\rm Mpc}$ \\
$\epsilon'$   & -                                              & ${0.43}$                       \\
$\mu_{\rm G}'$ & -                                              & ${0.24}$                     \\
$\mu_{\rm L}'$ & -                                              & ${0.55}$                      \\
\hline
Minimum $\chi^2$/d.o.f. & ${58/44}$                                   & ${57/41}$              \\
\hline\hline
\end{tabular}
\renewcommand{\arraystretch}{1.}
\end{center}
\end{table}

\clearpage
\clearpage

Non-thermal pressure possibly caused by turbulent gas and bulk motion 
causes a systematic error when comparing observations of clusters.
The authors estimated the fraction of non-thermal pressure
in the Coma Cluster \cite{Schuecker}, which can be larger than that of the thermal
pressure by $10$ percent.  
Here, we estimate how non-thermal pressure affects our fitting based on 
an estimation with a numerical simulation.
To this end, we estimate the hydrostatic masses by comparison with 
the X-ray brightness and SZ effect profiles of the Coma Cluster.
Here we define the non-thermal fraction $f_{\rm nth}$ by
$f_{\rm nth}\equiv{P_{\rm nth}}/({P_{\rm nth}+P_{\rm th}}),$
where $P_{\rm nth}$ and $P_{\rm th}$ are the non-thermal pressure
and the thermal pressure, respectively.
In the case including non-thermal pressure, 
the thermal pressure is replaced by $P_{\rm th}=(1-f_{\rm nth})P_{\rm tot}$.
We consider the following non-thermal pressure fraction as a function of the radius,
\begin{align}\label{nth}
  f_{\rm nth}(r)=\alpha_{\rm nt}(1+z)^{\beta_{\rm nt}}\left(\frac{r}{r_{500}}\right)
  ^{n_{\rm nt}}
  \left(\frac{M_{200}}{3\times 10^{14}M_{\odot}}\right)^{n_{\rm M}},
\end{align}
which is a theoretical prediction with numerical simulations in Ref.~\cite{Shaw,Battaglia}. 
$r_{500}$ and $M_{200}$ mean
  the radius and mass at the radius where the matter density in the galaxy cluster
  is 500 and 200 times of the critical density, respectively.
In the present paper we adopt $(\alpha_{\rm nt},~\beta_{\rm nt},~n_{\rm nt},
n_{\rm M})=(0.18,0.5,0.8.0.2)$,
which are the best-fit values in \cite{Battaglia} consistent with those in 
\cite{Schuecker}.

\begin{figure}[t]
\begin{center}
   \includegraphics[width=0.7\hsize]{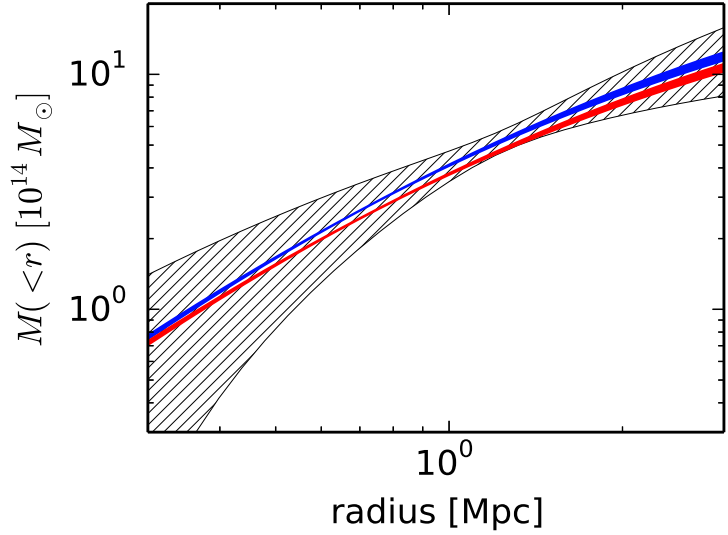}
\end{center}
  \caption{
    Spherical masses enclosed within the radii.
     The gray-hatched region denotes the $1\sigma$ uncertainty interval for
 the lensing mass determined solely by weak-lensing analysis \cite{Coma_WL1}.
 The blue-solid and red-solid regions denote 
    the $1\sigma$ uncertainty intervals for hydrostatic masses
 without and with the non-thermal pressure component, respectively, determined by our
 joint-fit method.
    The hydrostatic and lensing masses agree with each other,
    irrespective of the presence of a non-thermal pressure component.
\label{fig:mass_NT}
  }
\end{figure}

The best-fit profile in the presence of non-thermal pressure is not significantly 
altered, compared with the best-fit profile in the absence of non-thermal pressure.
Figure~\ref{fig:mass_NT} shows the enclosed mass profiles as a function of radius.
The gray-hatched region is the 1$\sigma$ uncertainty interval for the lensing mass.
The blue-solid and red-solid regions show the 1$\sigma$ uncertainty
intervals for hydrostatic masses fitted without and with 
non-thermal pressure, 
respectively. The hydrostatic mass estimates are in good agreement with
the lensing mass, regardless of the inclusion of the non-thermal pressure components.
This shows that our fitting method is not affected by non-thermal pressure,
so we do not consider the non-thermal effect when putting a constraint on the
modified gravity in the next section.

\section{Testing the generalized Galileon gravity model} 
\label{sec:GG}
We here consider the generalized cubic Galileon model, 
with action 
given by \cite{DKT},
\begin{align}
S=\int d^4x\sqrt{-g}\Bigl[G_4(\phi)R+K(\phi,X)-G_3(\phi,X)\square\phi
+{\cal L}_{\rm m}\Bigr],
\end{align}
where $K(\phi,X)$, $G_3(\phi,X)$ and $G_4(\phi)$ are arbitrary functions depending on 
the scalar field $\phi$
and its kinetic term $X\equiv -(\partial\phi)^2/2$ and ${\cal L_{\rm m}}$ is the matter Lagrangian. 
This model is a non-minimal coupling version of the kinetic gravity braiding mode \cite{Kimura2}, 
and a subclass of the most general second-order scalar-tensor theory 
\cite{Horndeski,Deffayet5,Kobayashi2} with
$G_{4X}=G_5=0$, where $G_{4X}\equiv\partial G_4/\partial X$.
The simplest cubic Galileon model is the case with 
$K=-X$, $G_3\propto X$, and $G_4=M_{\rm Pl}^2/2$, where $M_{\rm Pl}^2=1/(8\pi G)$ is the Planck mass.
  The DGP model is originally a 5-dimensional brane-world model, however, it can be effectively 
  described as a Galileon model.
Note that the DGP model has two branches of cosmological solutions,
the self-accelerating branch (sDGP) model \cite{Koyama1,Koyama2}
and the normal branch DGP (nDGP) model \cite{Deffayet4}.
The relation between the generalized Galileon model and the specific models 
are summarized in the appendix.

We consider perturbations of space-time metric and scalar field.
We choose the Newtonian gauge for the space-time metric~(\ref{metric}).
Assuming spherical symmetry of the system, within the sub-horizon scale 
with the quasi-static approximation
keeping with the Vainshtein feature,
the equations for the gravity and the scalar field lead to \cite{Kimura}
\begin{align}\label{force1}
\frac{d\Psi}{dr}&=\frac{GM(<r)}{r^2}-(\alpha+\xi)\frac{dQ}{dr}, \\
\label{force2}
\frac{d\Phi}{dr}&=-\frac{GM(<r)}{r^2}+\xi\frac{dQ}{dr}, \\
\label{force3}
\frac{dQ}{dr}&=\frac{r}{4\lambda^2}\left(1-\sqrt{1+\frac{8G\lambda^2\zeta M(<r)}{r^3}}
\right), 
\end{align}
where $Q({\bf x})$ is perturbation of the scalar field defined by
$\phi(t,{\bf x})=\phi(t)(1+Q({\bf x}))$, and 
$M(<r)\equiv 4\pi\int_0^rdr^\prime {r^{\prime}}^2\rho(r^\prime)$ is the 
enclosed mass of the 
halo within the physical radius $r$. 
Note that the perturbed values $\Psi$,
$\Phi$ and $Q$ in ({\ref{force1}})$\sim$({\ref{force3}}) are written
in the physical coordinate.
In ({\ref{force1}})$\sim$({\ref{force3}}), we introduce free model parameters 
$\alpha$, $\xi$, $\zeta$ and $\lambda^2$, which are determined by the arbitrary functions
$K$, $G_3$ and $G_4$. The expressions for 
$\alpha$, $\xi$, $\zeta$ and $\lambda^2$ are given in the appendix \ref{sec:coefficients}. 
The explicit expressions for the simplest cubic
Galileon, the sDGP and the nDGP models are also presented there.

Here, we define the Vainshtein radius $r_{\rm V}$ as
\begin{align}
r_{\rm V}\equiv [8G\lambda^2\zeta M_{\rm vir}]^{1/3}=\bigg[{8G\epsilon ^2 M_{\rm vir}\over H_0^2}\biggr]^{1/3},
\end{align}
where we define $\epsilon=\sqrt{H_0^2\lambda^2\zeta}$ using the Hubble constant $H_0$.
For $r\ll r_{\rm V}$, the scalar field can be negligible compared with the 
Newton potential, so Newtonian gravity is recovered.
For $r\gg r_{\rm V}$ the scalar field cannot be negligible, and we have 
\begin{align}\label{force1b}
\frac{d\Psi}{dr}&\simeq\frac{(1+\zeta(\alpha+\xi))GM(<r)}{r^2},\\
\label{force2b}
\frac{d\Phi}{dr}&\simeq -\frac{(1+\zeta\xi)GM(<r)}{r^2}.
\end{align}
Thus the gravitational and curvature potentials are modified
at $r\gg r_{\rm V}$.
These modifications affect both the gas and weak-lensing profiles.

We next construct observational quantities of 
the gas and weak-lensing profiles considering the scalar field.
Since gas components feel gravitational force through the gravitational 
potential $\Psi$, the X-ray brightness and the SZ profiles are
modified through modification of $\Psi$. 
On the other hand, the gravitational lensing is characterized by
the lensing potential
$(\Phi-\Psi)/2$, so the modified lensing potential alters
the observed lensing profile. We therefore introduce the parameters 
\begin{eqnarray}
&& \mu_{\rm G}\equiv (\alpha+\xi)\zeta,
\\
&& \mu_{\rm L}\equiv {1\over 2}(\alpha+2\xi)\zeta, 
\end{eqnarray}
with which we can write $d\Psi/dr\simeq(1+\mu_{\rm G})GM(<r)/r^2$ and  $d(\Psi-\Phi)/dr/2\simeq(1+\mu_{\rm L})GM(<r)/r^2$
at $r\gg r_{\rm V}$.

In the generalized Galileon model, with the use of parameters $\mu_{\rm G}$, $\mu_{\rm L}$ and $\epsilon$
our modeling for the electron pressure profile (\ref{pressureB1}) 
and the weak-lensing profile (\ref{kappa3}) are modified as follows:

\begin{align}\label{pressureB2}
&  P_{\rm e}(r)=P_0\exp\left(\int_0^rdr \frac{\mu m_{\rm p}}{kT_{\rm e}(r)}
\left[-\frac{GM(<r)}{r^2}+\frac{\mu_{\rm G}}{4\epsilon^2}H_0^2r
\left(1-\sqrt{1+12\epsilon^2\frac{\rho_{\rm s}}{\rho_{\rm c0}}\frac{r_{\rm s}^3}{r^3}m(r)}\right)
\right]\right),
\\
&\kappa(r_\perp)= 
\frac{2}{\Sigma_{\rm c}}
\int_0^{\infty}dz\biggl[\rho(r)-\frac{\mu_{\rm L}\rho_{\rm c0}}{2\epsilon^2}
\left(1-\sqrt{1+12\epsilon^2
\frac{\rho_{\rm s}}{\rho_{\rm c0}}\frac{r_{\rm s}^3}{r^3}m(r)}\right)
\nonumber\\
&~~~~~~~~~~~~~~~~~~~~~~~~~~~~~~~~~~~~~~~~~~~~~~~~
+\frac{\rho(r)-3\rho_{\rm s}r_{\rm s}^3m(r)/r^3}
{\sqrt{1+12\epsilon^2\rho_{\rm s}r_{\rm s}^3m(r)/\rho_{\rm c0}r^3}}\mu_{\rm L}\biggr].
\label{kappa4}
\end{align}
Since the gas pressure tracing the matter density 
deceases with the cluster-centric radius increasing, 
the pressure gradient is restricted to $dP_{\rm e}/dr<0$. This gives the
constraints on $\mu_{\rm G}$.

\begin{figure}[t]
    \includegraphics[width=\hsize]{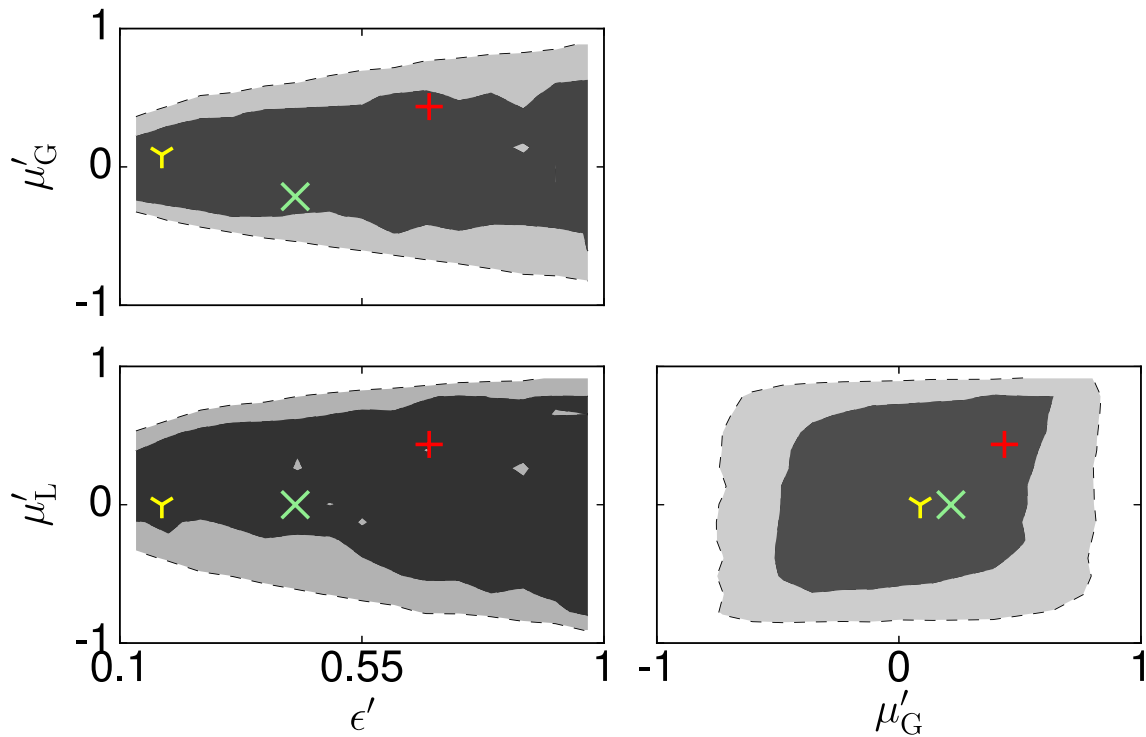}
  \caption{
 The 68\% (dark gray regions) and the 95\% CL (light gray regions) 
    2-dimensional marginalized contours for the generalized Galileon model parameters. 
The red-plus, green-cross and yellow-triangle points present models for the simplest cubic
 Galileon, the sDGP and the nDGP models, respectively.}
    \label{fig:constraint}
\end{figure}

Instead of $\mu_{\rm G}$, $\mu_{\rm L}$ and $\epsilon$, we introduce 
$\mu_{\rm G}^\prime={\mu_{\rm G}}/({1+|\mu_{\rm G}|})$,~~ 
$\mu_{\rm L}^\prime={\mu_{\rm L}}/({1+|\mu_{\rm L}|})$, and 
$\epsilon^\prime=1-\exp(-\epsilon)$,
which span the complete available parameter space of $\mu_{\rm G}^\prime$ 
and $\mu_{\rm L}^\prime$ in the interval $[-1,1]$ and $\epsilon^\prime$ in 
the interval $[0,1]$, respectively.
General relativity is recovered when $\mu_{\rm G}'=\mu_{\rm L}'=0$ or 
$\epsilon'\to 1$. %
Using the same method adopted for Newtonian case, we perform an MCMC analysis
for the modified gravity model
including $8$ parameters with the chi-squared $\chi^2_{\rm BX+SZ+WL}$, defined by (\ref{chi2}).

Figure~\ref{fig:mcmc} shows the 2-dimensional marginalized contours of the
different combinations between the model parameters.
The best-fit parameters and their 1-dimensional marginalized 68\% errors
are listed in the Table.~\ref{table1}. The red curve 
in each panel of figure~\ref{fig:observed2_MG} shows the best-fit profile for 
the generalized Galileon model with the minimum value of the chi-squared/d.o.f.,
$\chi^2_{\rm XB+SZ+WL}/{\rm d.o.f.}=57/41$.
These profiles almost overlap with the profiles for 
Newtonian gravity (black dashed curves), which shows that the large deviation 
from Newtonian gravity is rejected.
We note that there is no significant difference between the red and black curves in the 
  best-fit profiles. There is a slight difference in the shear profiles at the 
  large radius $r>1$Mpc, which seems to be originated from the large errorbars of the 
  shear data.

Figure \ref{fig:constraint} shows 2-dimensional marginalized contours
of the confidence levels for the parameters $\mu_{\rm G}'$, $\mu_{\rm L}'$ and 
$\epsilon'$. $\mu_{\rm G}'$ and $\mu_{\rm L}'$ are parameters from
the modification of the gravitational potential and the lensing potential,
and $\epsilon'$ is a parameter characterizing the Vainshtein radius.
Large values of $\mu_{\rm G}$ and $\mu_{\rm L}$ are rejected at the $68 \%$ confidence level, 
  which indicates that the possibility of a large deviation from the Newtonian gravity is ruled out,
  depending on the parameter $\epsilon$.
When $\epsilon$ is smaller, the Vainshtein radius becomes smaller, we can put a tighter constraint
on $\mu_{\rm G}$ and $\mu_{\rm L}$. However, $\epsilon$ is large, the Vainshtein radius becomes large, 
which makes difficult to distinguish between the Newtonian gravity model and the modified gravity model
due to the Vainshtein mechanism.
The red-plus, green-cross and yellow-triangle points in figure~\ref{fig:constraint} show 
the representative models, the simplest cubic Galileon model,
the sDGP model and nDGP model, respectively, at the redshift $z=0.0236$.
The parameter values for each models are shown in Table~\ref{table2}.

In a previous work \cite{Narikawa},
a constraint only on the parameter space 
$\mu_{\rm L}$ and $\epsilon$ is obtained, based on the lensing observations.
As a other recent related work, Barreira et al. investigated cluster masses 
and the concentration parameters in modified gravity models 
from shear profiles \cite{Barreira2}. They focused their investigation on the mass-concentration 
relation of 19 X-ray selected clusters from the CLASH survey in the simplest cubic 
Galileon and Nonlocal gravity models. 
They found that the mass-concentration relation obtaining from the shear profiles 
for the cubic Galileon model is the same as those for the $\Lambda$CDM model, 
but no stringent constraint on the modified gravity models is obtained.
Unfortunately the constraint obtained in the present paper is not very stringent too,
but one can find the following possibility.
We emphasize that models with $\mu_{\rm L}=0$
like the sDGP and the nDGP models are indistinguishable with
Newtonian gravity in the method based on the lensing observations.
On the other hand, our method of combining 
the gas and weak-lensing
profiles can solve the problem from this degeneracy.
Future observations would improve the constraint.

\begin{table}[t]
\caption{
  Values of modified gravity parameters for each model at the redshift $z=0.0236$.
\label{table2}
}

\begin{center}
\renewcommand{\arraystretch}{1.5}
\begin{tabular}{c|c|c|c}
\hline\hline
~~~~Models~~~~ & ~~~~$\epsilon~(\epsilon')$~~~~ & ~~~~$\mu_{\rm G}~(\mu_{\rm G}')$~~~~ & ~~~~$\mu_{\rm L}~(\mu_{\rm L}')$~~~~ 
\\
\hline
simplest cubic Galileon  & ${0.77~(0.44)}$    & ${0.77~(0.44)}$    & ${1.12~(0.67)}$          \\
sDGP               & ${-0.26~(-0.22)}$  & ${0~(0)}$          & ${0.53~(0.43)}$          \\
nDGP               & ${0.20~(0.18)}$    & ${0~(0)}$          & ${0.10~(0.09)}$           \\
\hline\hline
\end{tabular}
\renewcommand{\arraystretch}{1.}
\end{center}
\end{table}

\begin{table}[t]
\caption{
  Same as Table 1 but for the results of the generalized Galileon model in the unscreened limit 
    with only fixing $\epsilon=0$ (second column), and the case with fixing all 
    the modified gravity parameters $\epsilon'=0.05$, $\mu_{\rm G}'=0.2$, and $\mu_{\rm L}'=0$ (third column).
\label{table3}}
\begin{center}
  \renewcommand{\arraystretch}{1.5}
\begin{tabular}{c|c|c}
\hline\hline
Parameter & Modified gravity (unscreened) & Modified gravity (fifth force) 
\\
\hline
$M_{\rm vir}$  & ${1.26}^{+0.15}_{-3.85}\times 10^{15}~M_\odot$ & ${0.86}^{+0.05}_{-0.05}\times 10^{15}~M_\odot$\\ 
$c$           & ${3.78}^{+0.13}_{-0.55}$ & ${3.84}^{+0.24}_{-0.28}$\\
$n_0$         & ${6.15}^{+0.27}_{-0.29}\times 10^{-3}{\rm /cm^3}$ & ${6.20}^{+0.27}_{-0.32}\times 10^{-3}{\rm /cm^3}$\\
$T_0$         & ${6.35}^{+0.13}_{-0.11} ~{\rm keV}$ & ${6.36}^{+0.12}_{-0.12} ~{\rm keV}$\\
$b_1$         & ${2.6}$ (fixed) & ${2.6}$ (fixed)\\
$b_2$         & ${0.5}$ (fixed) & ${0.5}$ (fixed)\\
$r_1$         & ${0.75}^{+0.06}_{-0.07} ~{\rm Mpc}$ & ${0.75}^{+0.06}_{-0.06} ~{\rm Mpc}$\\
$\epsilon'$   & ${0}$ (fixed)              & ${0.05}$ (fixed)                      \\
$\mu_{\rm G}'$ & ${-0.10}$                 & ${0.2}$ (fixed)                        \\
$\mu_{\rm L}'$ & ${-0.05}$                 & ${0}$  (fixed)                        \\
\hline
Minimum $\chi^2$/d.o.f.& ${57/42}$               & ${60/44}$                      \\    
\hline\hline
\end{tabular}
\renewcommand{\arraystretch}{1.}
\end{center}

\end{table}

\section{Discussion}
\label{sec:Discussion}
\subsection{Degeneracies on parameters}
On the MCMC analysis in the previous section, we do not take 
the range of $\epsilon$, $[0,0.1]$,
into account because it is hard to converge the MCMC runs because of degeneracy in the 
parameter space.
Here, we treat this parameter region for complemental discussion.

First, taking the limit $\epsilon\to 0$, which means the fifth force is unscreened everywhere,
the solutions of the gas pressure (\ref{pressureB2}) and the convergence (\ref{kappa4})
are reduced to
\begin{align}\label{pressureB3}
&  P_{\rm e}(r)=P_0\exp\left(\int_0^rdr \frac{\mu m_{\rm p}}{kT_{\rm e}(r)}
\left[-\frac{GM(<r)}{r^2}(1+\mu_{\rm G})\right]\right),
\\
&\kappa(r_\perp)= (1+\mu_{\rm L})\frac{2}{\Sigma_{\rm c}}
\int_0^{\infty}dz\rho(r).
\label{kappa5}
\end{align}
Then, the pressure profile and the convergence profile are simply modified by a factor of
$(1+\mu_{\rm G})$ and $(1+\mu_{\rm L})$, respectively.
In this case, we have $P_{\rm e}\propto (1+\mu_{\rm G})M_{\rm vir}m(c)/c^3$ and
$\kappa\propto \rho_{\rm s}\propto (1+\mu_{\rm L})M_{\rm vir}c^3/m(c)$, then, there are degeneracies
between the parameters, $M_{\rm vir}, c, \mu_{\rm G}$, and $\mu_{\rm L}$.
Figure \ref{fig:mcmc2} compares the results of the MCMC analysis with fixing 
$\epsilon=0$ (dark blue region ($68$\% CL) and mid blue region ($95$\% CL)) 
and the results of the Newtonian gravity (dark gray region and mid gray region), 
which is the same as those of figure \ref{fig:mcmc0}. 
The best-fit parameters are shown in the Table.~\ref{table3}.
The CL contours of the blue regions reflect the degeneracy between the 
parameters, $M_{\rm vir}, c, \mu_{\rm G}$, and $\mu_{\rm L}$.

Next, we show how the presence of the fifth force affects the parameter estimation.
For example, the blue confidence contours in figure \ref{fig:mcmc3} shows the 
$68$\% and $95$\% confidence contours of the case 
with fixing $\epsilon=0.05$, $\mu_{\rm G}=0.2$ and $\mu_{\rm L}=0$. 
$M_{\rm vir}$ and $c$ are different from those of the Newtonian gravity (gray regions), 
but other parameters, $n_0$, $T_0$ and $b_1$, are not changed.
The minimum value of the chi-squared/d.o.f. in the presence
of the fifth force is $\chi^2_{\rm XB+SZ+WL}/{\rm d.o.f.}=60/44$, which
is almost the same as the Newtonian case, despite the different cluster
parameter, $M_{\rm vir}\sim 0.9\times 10^{15}M_\odot$
(see Table.~\ref{table3}).
This result
exemplifies that the presence of the attractive fifth force
affects the estimation of the NFW parameters, $M_{\rm vir}$ and $c$.
This is understood as the consequence of the degeneracy between the modification 
parameters $\mu_{\rm G}$ and $\mu_{\rm L}$ and $M_{\rm vir}$ and $c$.

\subsection{Systematic errors}
We shall
discuss possible systematic errors. In our analysis, we have 
assumed spherical symmetry for the matter distribution and an
equilibrium state for the gas component of the balance between the 
pressure gradient and the gravitational force and the fifth force 
in the case of its presence. We have demonstrated that 
non-thermal pressure at the level suggested by 
numerical simulations does not alter our results.
A future X-ray satellite, ASTRO-H \cite{ASTROH}, will observe turbulent gas motion in the Coma Cluster
  in more detail, which will be informative regarding our result.
However, observations of the Coma Cluster suggest substructures 
\cite{Coma_WL1,Coma_WL2,Honda,Watanabe,Arnaud,Neumann} and orientation dependence
\cite{Coma_xray3,Coma_xray4,Gastaldello}, so the Coma Cluster is not thought to 
be a relaxed system. Dynamical states of the Coma Cluster would 
give a systematic difference between our results and temperature measurement.
Our fitting results show that the temperature of the Coma Cluster
is around $6.4$ keV (see Table.~\ref{table1}), but this result
seems lower than those of X-ray observations \cite{Coma_xray1,Coma_xray2,Coma_xray3,Coma_xray4,Coma_XT1,Coma_XT2,Gastaldello}, which estimate that
the temperature of the Coma Cluster is around $8$--$9$ keV.
Comparing the mass-temperature scaling relation for a sample of
relaxed clusters \cite{Okabe2} with an X-ray temperature observation of the Coma Cluster
\cite{Coma_xray1,Coma_xray2},
the observed temperature is higher that the
temperature expected by the mass. The enhancement is at 3$\sigma$ level of intrinsic scatter \cite{Okabe2}.
Similar results of high temperatures have also been reported by a comparison
with other clusters \cite{Pimbblet}. 
Depending on the orientation and excluding the central region,
  the temperature of the Coma Cluster could be around $6$--$7$ keV \cite{Coma_xray1,Coma_xray3},
  but it is difficult to take this dependence into account.
Therefore, systematic error of temperature in the Coma Cluster
would cause a substantial influence to the proposed fitting method.  
In order to reduce a possible dependence of cluster-dynamical
states and halo triaxiality, it is of vital importance to increase the
number of sampling clusters. Ongoing and future multi-wavelength
surveys such as the Hyper Suprime-Cam (HSC) optical survey\footnote{http://subarutelescope.org/Projects/HSC/surveyplan.html},
the Dark Energy Survey
(DES) \cite{DES},  the {\it eROSITA} X-ray survey \cite{eROSITA}, and the ACT-Pol \cite{ACTPol} and SPT surveys \cite{SPT} will
be powerful aids to better constraining the gravity model.

\section{Summary and Conclusion} 
\label{sec:summary}

In this paper, we obtained a constraint on the generalized
Galileon model through the Coma Cluster observations of X-ray brightness,
the SZ effect and weak lensing. We have constructed a simple analytic model of 
the gas distribution profiles and the weak-lensing profile (c.f. \cite{Terukina2,Harry,Narikawa}).
The fifth force affects not only the gas distribution but also 
the weak-lensing profile. In general, the effects depend on different parameters 
characterizing the generalized Galileon model. 
These features can be investigated by combination of the observations of a 
galaxy cluster reflecting the gas density profile and the lensing signals. 
Their multi-wavelength observations are complementary to each other, 
and are useful to put a constraint on the 
modified gravity model by breaking the degeneracy between the model parameters. 
Systematic study compiling  multi-wavelength datasets for a large number of clusters
enables us to well reduce the systematic errors and
constraints on the modified gravity models. 
However, the degeneracy between the parameters,
 $M_{\rm vir}$, $c$, $\mu_{\rm G}$ and $\mu_{\rm L}$, persists in the limit of 
the weak screening of the fifth force, which affects the estimation of the cluster parameters.
Future and ongoing
surveys and their joint analysis would be a powerful aid to obtaining a more stringent constraint on modified gravity models.

\section*{Acknowledgment}
This work is supported by a research support program of Hiroshima
University, the Funds for the Development of Human Resources in Science and Technology under MEXT, Japan, and 
 Core Research for Energetic Universe at Hiroshima University (the MEXT
 Program for Promoting the Enhancement of Research Universities, Japan).
We thank Kuniaki Masai, Lucas Lombriser, Yuying Zhang and Tatsuya Narikawa for their useful comments and discussions. 
N. Okabe is supported by a Grant-in-Aid from the Ministry of Education, Culture, 
Sports, Science, and Technology of Japan (26800097).



\appendix
\section{Definitions of the coefficients}
\label{sec:coefficients}
In this appendix, we summarize the coefficients 
between the generalized Galileon model and the specific models used in section~\ref{sec:GG}
 (see also \cite{Narikawa,Narikawa3}). 
The coefficients in the perturbation equations (\ref{force1})--(\ref{force3}) 
are defined as 
\begin{align}
\alpha&=\alpha_1, \\
\xi&=\alpha_2, \\
\zeta&=\frac{4(\alpha_1+\alpha_2)}{\beta}\frac{G_4H}{\dot\phi\phi}, \\
\lambda^2&=\frac{2\beta_0G_4\phi H}{\beta X\dot\phi}, \\
\beta&=-4(\alpha_0+2\alpha_1\alpha_2+\alpha_2^2)\frac{G_4H^2}{\dot\phi^2},
\end{align}
where
\begin{align}
\alpha_0&=\left(\frac{\dot \Theta}{H^2}+\frac{\Theta}{H}-2G_4-4\frac{\dot G_4}{H}
-\frac{{\cal E+P}}{2H^2}\right){1\over 2G_4}, \\
\alpha_1&=\left(2\frac{\dot\phi G_{4\phi}}{H}\right){1\over 2G_4}, \\
\alpha_2&=\left(\frac{\dot\phi XG_{3X}}{H}-\frac{\dot\phi G_{4\phi}}{H}\right){1\over 2G_4}, \\
\beta_0&=\left(\frac{\dot\phi XG_{3X}}{H}\right){1\over 2G_4}, \\
\Theta&=-\dot\phi XG_{3X}+2HG_4+\dot\phi G_{4\phi}.
\end{align}
These coefficients are determined by the background solution, which follows:
\begin{align}
&2XK_X-K+6X\dot\phi HG_{3X}-2XG_{3\phi}-6H^2G_4-6H\dot\phi G_{\phi}=\rho_{\rm m},\\
&K-2X\left(G_{3\phi}+\ddot\phi G_{3X}\right)+2\left(3H^2+2\dot H\right)G_4
+2\left(\ddot\phi+2H\dot\phi\right)G_{4\phi}+4XG_{4\phi\phi}=0,
\end{align}
where $\rho_{\rm m}$ is the non-relativistic matter energy density and
$H=\dot a/a$ is the Hubble parameter.
The background equation for the scalar field is written as
\begin{align}
\dot J +3HJ-P=0,
\end{align}
with
\begin{align}
J&\equiv \dot\phi K_X+6HXG_{3X}-2\dot\phi G_{3\phi},\\
P&\equiv K_\phi-2X\left(G_{3\phi\phi}+\ddot\phi G_{3\phi X}\right)
+6\left(2H^2+\dot H\right)G_{4\phi}.
\end{align}

The simplest cubic Galileon model is defined by 
$G_4=M_{\rm Pl}^2/2$, $K=-X$ and $G_3=(r_{\rm c}^2/M_{\rm Pl})X$,
which corresponds to taking $c_1=-1$ in Ref.~\cite{Barreira1},
and thus the coefficients in the perturbation equations are 
\begin{align}
\alpha&=0, \\
\xi&=4\pi G_3G_{3X}\dot\phi^2\phi, \\
\zeta&=\frac{G_{3X}\dot\phi^2}{\beta\phi}, \\
\lambda^2&=\frac{G_{3X}\phi}{\beta}, \\
\beta&=-1+2G_{3X}(\ddot\phi+2H\dot\phi)-4\pi G_3G_{3X}^2\dot\phi^4. 
\end{align}
When we adopt the late time de Sitter attractor solution \cite{Nesseris},
\begin{align}
\left(\frac{H(a)}{H_0}\right)^2=\frac{1}{2}\left[\frac{\Omega_{\rm m0}}{a^3}
+\sqrt{\left(\frac{\Omega_{\rm m0}}{a^3}\right)^2+4(1-\Omega_{\rm m0})}\right].
\end{align}
The combinations $\xi\zeta$ and $\lambda^2\zeta$ are given by
\begin{align}
\xi\zeta&=\frac{(1-\Omega_{\rm m})(2-\Omega_{\rm m})}
{\Omega_{\rm m}(5-\Omega_{\rm m})},\\
\lambda^2\zeta&=\left(\frac{2-\Omega_{\rm m}}
{H\Omega_{\rm m}(5-\Omega_{\rm m})}\right)^2,
\end{align}
where $\Omega_{\rm m}(a)=\rho_{\rm m}(a)/3M_{\rm Pl}^2H^2(a)$ is the matter density parameter.

Within the sub-horizon approximation, the DGP model \cite{Koyama1,Koyama2,Falck} 
can be effectively described by the coefficients
\begin{align}
\alpha&=-1, \\
\xi&=\frac{1}{2}, \\
\zeta&=-\frac{2}{3\beta}, \\
\lambda^2&=-\frac{r_{\rm c}^2}{3\beta}, \\
\beta&=1\pm 2Hr_{\rm c}\left(1+\frac{\dot H}{3H^2}\right),
\end{align}
where the sign ``$\pm$'' in $\beta$ represents the case of the sDGP model with
``$-$'' sign and the nDGP model with ``$+$'' sign.
For the sDGP model, we adopt the self-accelerating background solution, which is
specified by the modified Friedmann equation
in the sDGP model \cite{Deffayet4}, 
\begin{align}
\frac{H(a)}{H_0}=\frac{1-\Omega_{\rm m0}}{2}
+\sqrt{\frac{\Omega_{\rm m0}}{a^3}+\frac{(1-\Omega_{\rm m0})^2}{4}},
\end{align}
and $r_{\rm c}=1/(1-\Omega_{\rm m0})H_0$. 
On the other hand, the nDGP model has no self-accelerating solution 
without introducing the cosmological constant \cite{Deffayet1,Deffayet3}.
Here we consider the nDGP model with introducing a dynamical dark
energy component on the brane, which is tuned such that the background
evolves as in the lambda cold dark matter model \cite{FSchmidt}.

\begin{landscape}
\begin{figure}[t]
    \includegraphics[width=\hsize]{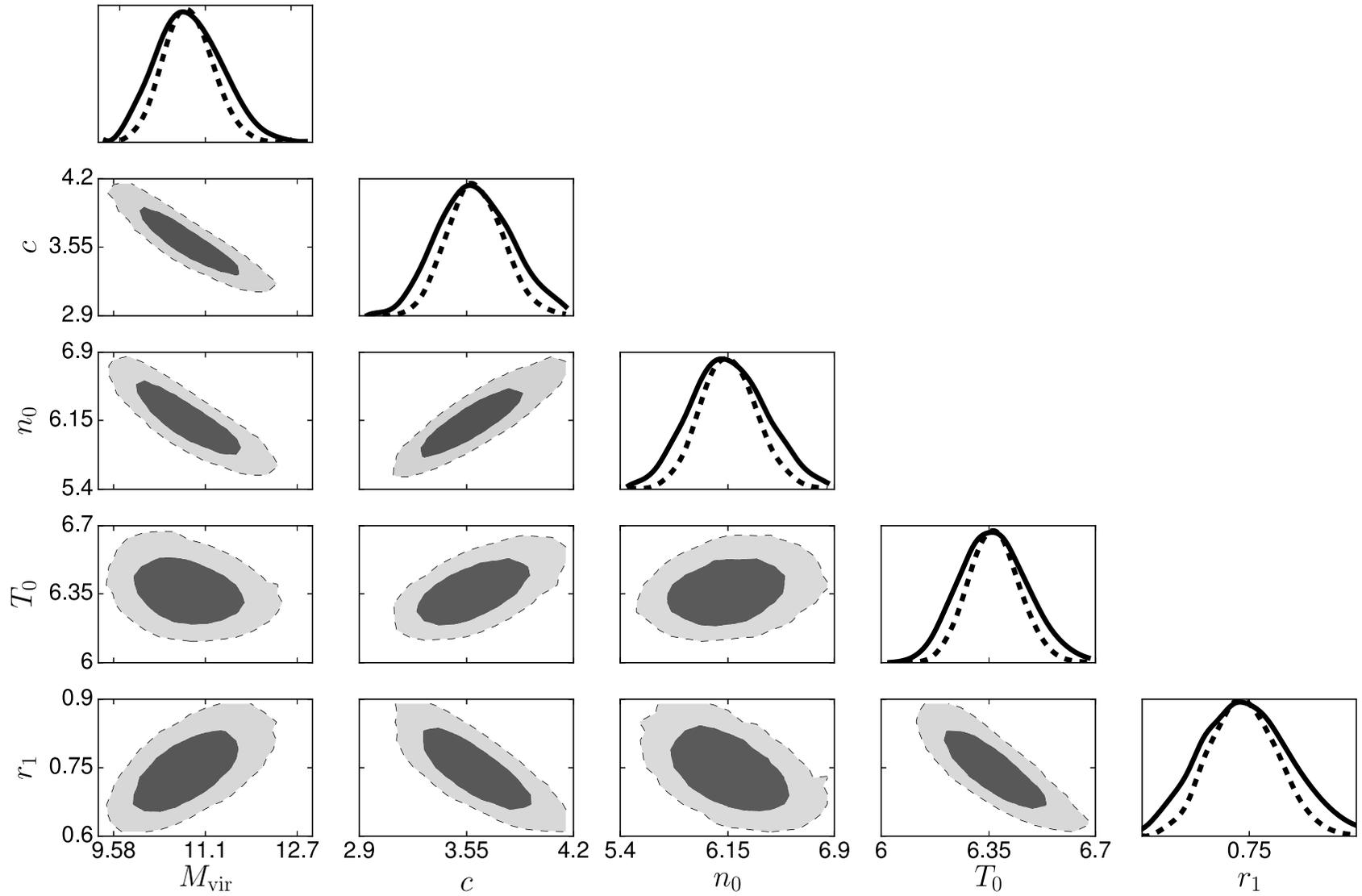}

  \caption{
The 68\% (dark gray region) and the 95\% CL (mid gray region) 
2-dimensional marginalized contours for the 5 model parameters,
$M_{\rm vir}~[10^{14}M_\odot]$, $c$, $n_0~[10^{-3}{\rm cm}^{-3}]$, $T_0~[{\rm keV}]$ and $r_1~[{\rm Mpc}]$.
The rightmost plots of each row show the 1-dimensional marginalized constraints (solid) and likelihood distributions (dotted).
    \label{fig:mcmc0}
  }
\end{figure}
\end{landscape}

\begin{landscape}
\begin{figure}[t]
    \includegraphics[width=\hsize]{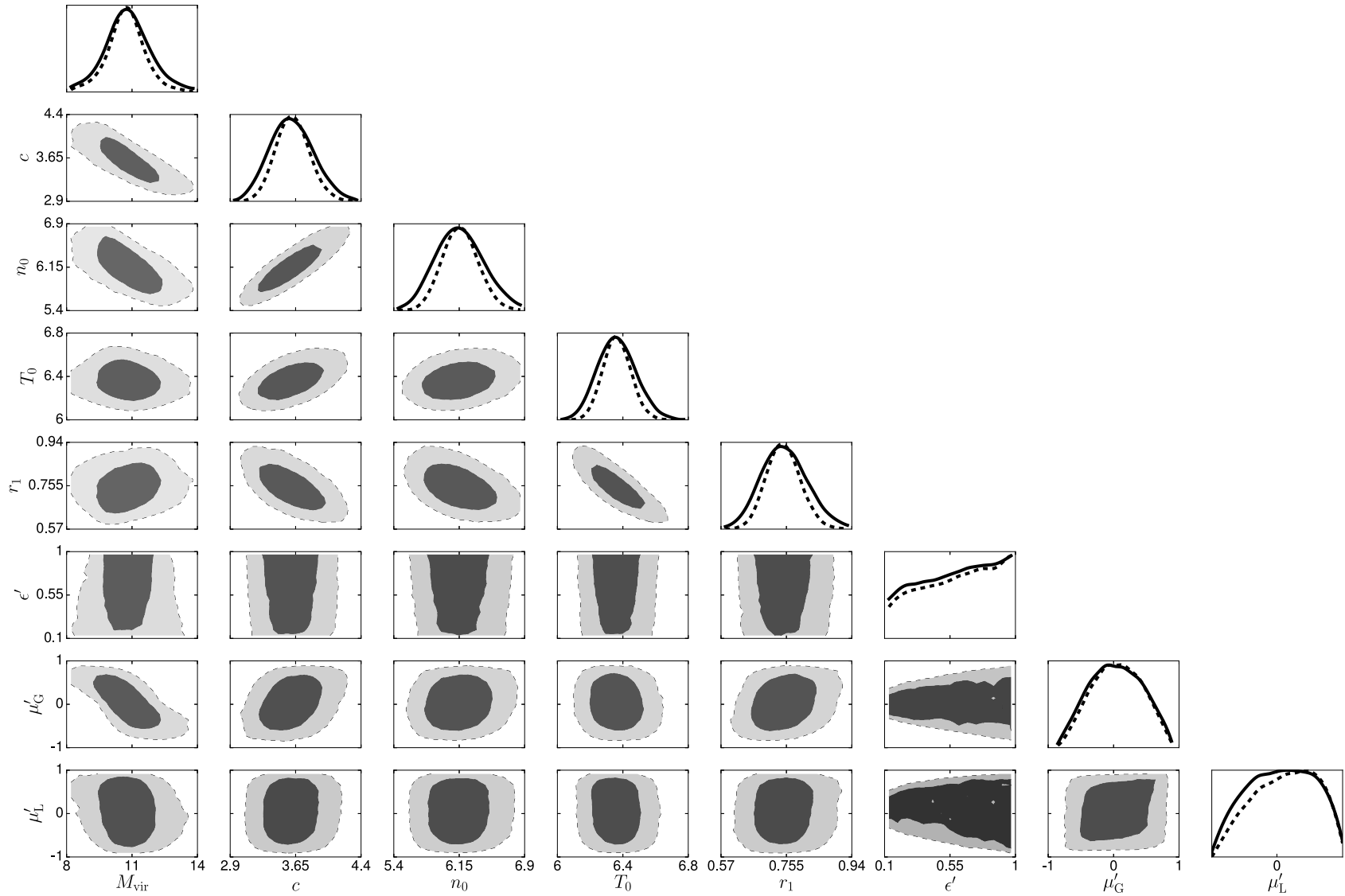}

  \caption{
    Same as figure \ref{fig:mcmc0} but when including the modification parameters 
    $\epsilon'$, $\mu_{\rm G}'$ and $\mu_{\rm L}'$ in the MCMC analysis.
    \label{fig:mcmc}
  }
\end{figure}
\end{landscape}

\begin{landscape}
\begin{figure}[t]
    \includegraphics[width=\hsize]{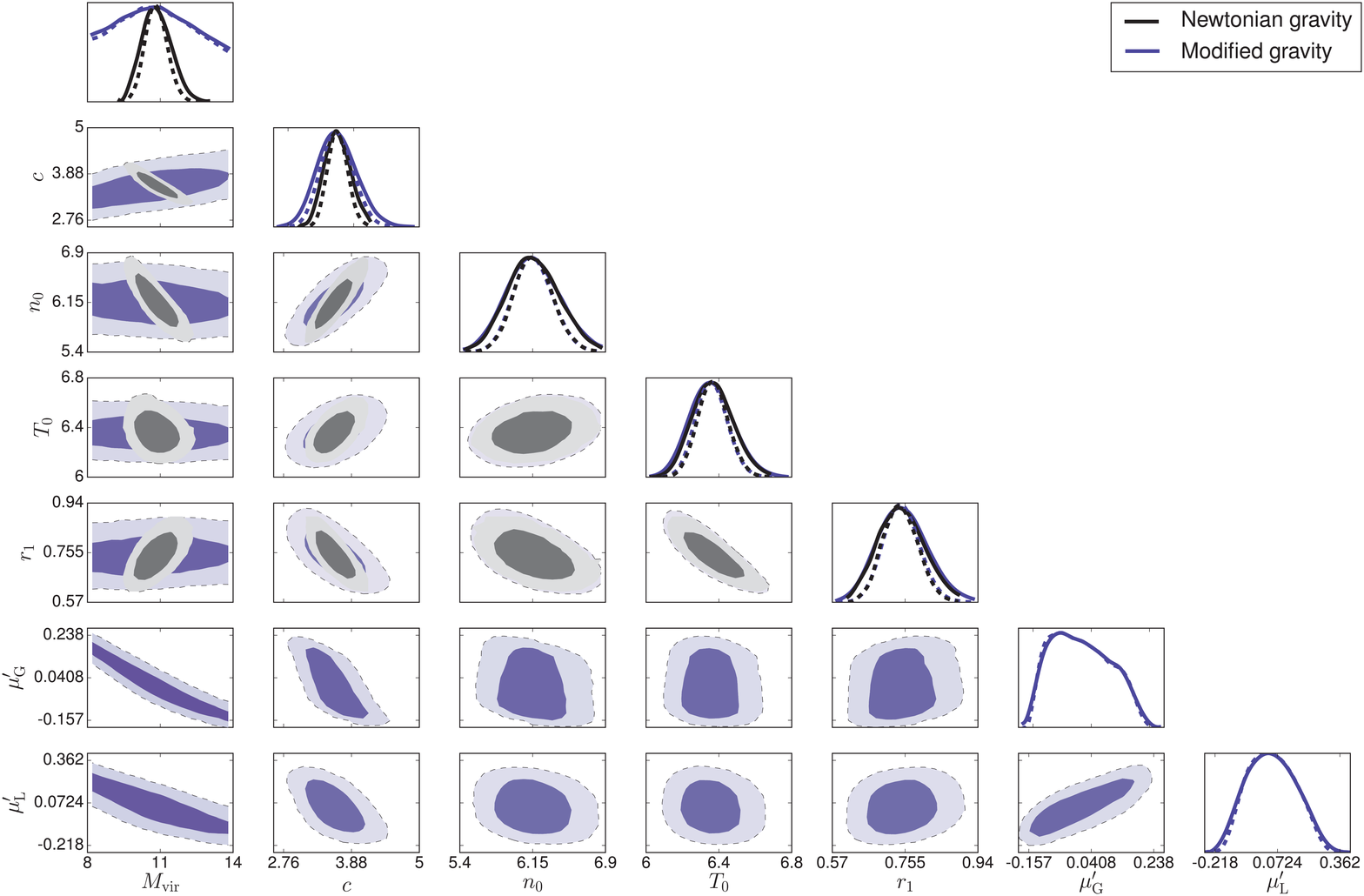}

  \caption{
The dark blue region ($68$\% CL) and the mid blue region ($95$\% CL)     
are the results of the MCMC analysis for the modified gravity model 
with fixing $\epsilon=0$. 
The dark gray region  ($68$\% CL) and the mid gray region ($95$\% CL)   
are the results for the Newtonian gravity (same as the figure 4).
    \label{fig:mcmc2}
  }
\end{figure}
\end{landscape}

\begin{landscape}
\begin{figure}[t]
    \includegraphics[width=\hsize]{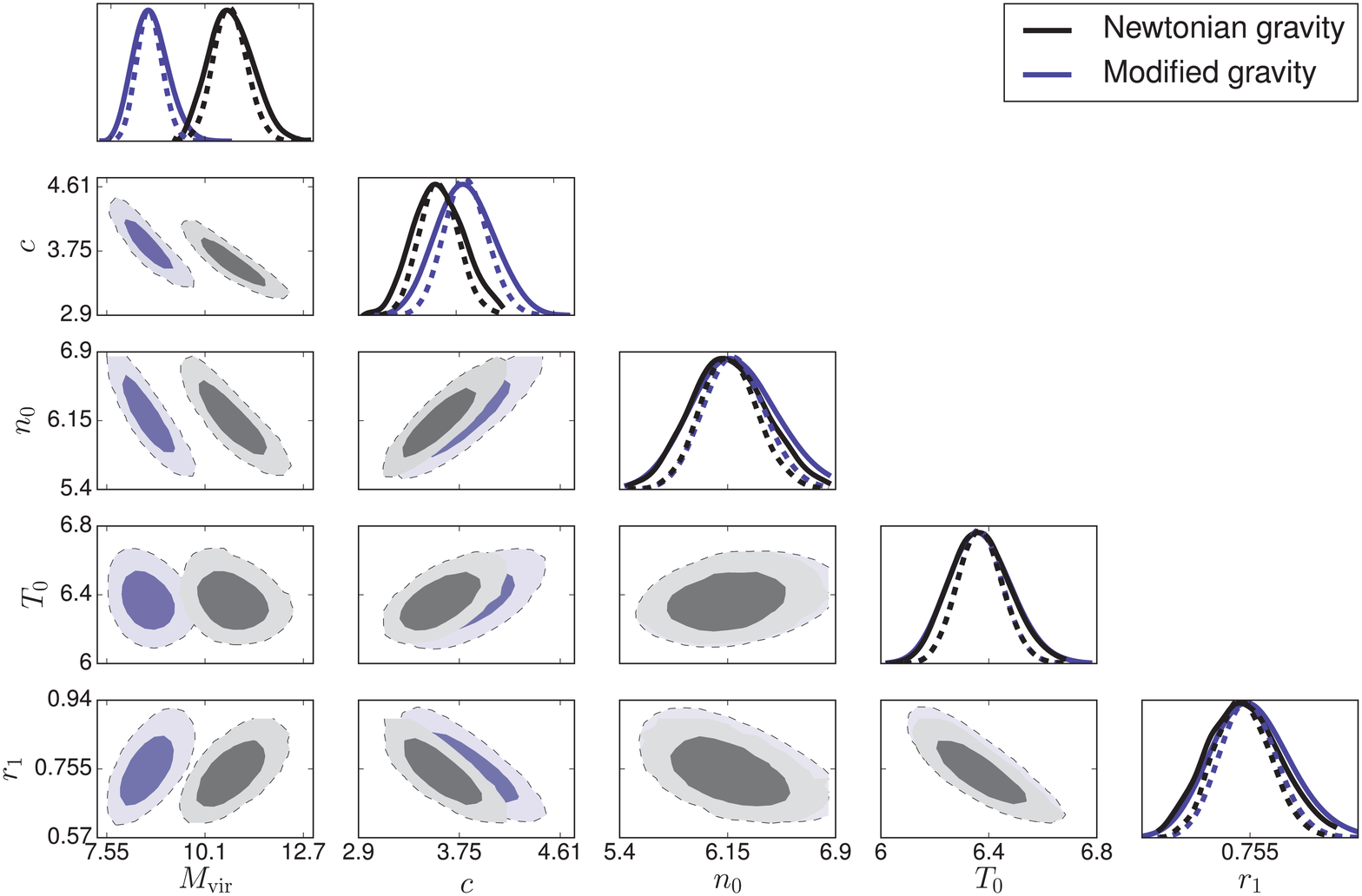}

    \caption{
The dark blue region ($68$\% CL) and the mid blue region ($95$\% CL)     
are the results of the MCMC analysis for the modified gravity model 
with fixing $\epsilon^\prime=0.05$, $\mu_{\rm G}^\prime=0.2$ and
$\mu_{\rm L}^\prime=0$. The dark gray region  ($68$\% CL) and the mid gray region ($95$\% CL)   
are the results for the Newtonian gravity (same as the figure 4).
    \label{fig:mcmc3}
  }
\end{figure}
\end{landscape}

\end{document}